\documentclass[aps,prb,nopacs,twocolumn,groupedaddress]{revtex4-1}

\usepackage{amsmath, amsthm, amssymb}
\usepackage{bm}
\usepackage{graphicx}
\usepackage{color}
\usepackage{hyperref}
\usepackage{cleveref}

\newcommand{\req}[1]{Eq.~(\ref{#1})}
\newcommand{\reqs}[1]{Eqs.~(\ref{#1})}

\newcommand{\kmpar}{\lambda_{\mathrm{SO}}}
\newcommand{\stpar}{\lambda_{\mathrm{v}}}
\renewcommand{\Re}{\rm{Re}}
\renewcommand{\Im}{\rm{Im}}

\begin{document}

%Title of paper
\title{Multiple odd-frequency superconducting states in buckled quantum spin Hall insulators with time-reversal symmetry}

\author{Dushko Kuzmanovski}
\affiliation{Department of Physics and Astronomy, Uppsala University, Box 516, SE-751 20 Uppsala, Sweden}

\author{Annica M.~Black-Schaffer}
\affiliation{Department of Physics and Astronomy, Uppsala University, Box 516, SE-751 20 Uppsala, Sweden}

\date{\today}

\begin{abstract}
We consider a buckled quantum spin Hall insulator (QSHI), such as silicene, proximity coupled to a conventional spin-singlet $s$-wave superconductor. Even limiting the discussion to the disorder-robust $s$-wave pairing symmetry, we find both odd-frequency ($\omega$) spin-singlet and spin-triplet pair amplitudes, both of which preserve time-reversal symmetry. Our results show that there are two unrelated mechanisms generating these different odd-$\omega$ pair amplitudes. The spin-singlet state is due to the strong interorbital processes present in the QSHI. It exists generically at the edges of the QSHI, but also in the bulk in the heavily doped regime if an electric field is applied. The spin-triplet state requires a finite gradient in the proximity-induced superconducting order along the edge, which we find is automatically generated at the atomic scale for armchair edges but not at zigzag edges. In combination these results make superconducting QSHIs a very exciting venue for investigating not only the existence of odd-$\omega$ superconductivity but also the interplay between different odd-$\omega$ states.
\end{abstract}

\maketitle

% ------------------------------------ %
% INTRODUCTION:
% ------------------------------------ %
\section{\label{sec:Intro}Introduction}
As was originally shown already by Berezinskii,~\cite{Berezinskii1974} the superconducting order parameter may be an odd function of relative time or, equivalently, frequency ($\omega$). This leads to an extension of the usual symmetry classification for superconducting states, allowing also for the possibility of spin-singlet odd-parity (\textit{p}, \textit{f}, \ldots) or spin-triplet even-parity (\textit{s}, \textit{d}, \ldots) states, without violating Fermi-Dirac statistics. 

While there exist proposals for a thermodynamically stable odd-$\omega$ superconducting order parameter,~\cite{Berezinskii1974, Kirckpatrick1991, AVB1, AVB2, Coleman1993} a more promising direction has been to induce odd-$\omega$ pair amplitudes in nonuniform systems, such as surfaces and interfaces with conventional even-$\omega$ superconducting order also present in at least part of the system. For example, superconductor-ferromagnet junctions break spin-rotation symmetry and have been shown to transform a conventional spin-singlet $s$-wave superconducting order parameter into an odd-$\omega$ spin-triplet (OT) $s$-wave pair amplitude with an unusually long-range decay in the ferromagnetic region.~\cite{Bergeret2005, Bergeret2001} Recent experimental advances have also been able to provide surmounting and diverse evidence of this odd-$\omega$ state in this type of junctions.~\cite{TcvsAngleExp, paraMeisExp, STMSigExp}
This idea has recently been extended to bulk systems with broken spin-rotation symmetry, where odd-$\omega$ spin-triplet $s$-wave pairing has also been shown to appear.~\cite{Triola2016}
Alternatively, nonmagnetic interfaces can induce odd-$\omega$ spin-singlet (OS) $p$-wave pairing,~\cite{Tanaka2007a, Tanaka2007b} due to translation symmetry breaking. However, the $p$-wave symmetry is notably less stable to disorder and therefore practically much less important.~\cite{Tanaka2007c}

Multiband, or equivalently multiorbital, superconductors have recently been shown to offer  another, very different, possibility for odd-$\omega$ superconductivity.~\cite{AMBS_Multiband2013, Komendova2015} Here the band label allows for additional symmetries for the superconducting state, such that the overall behavior under spin ($S$), spatial parity ($P$), orbital ($O$), and time-inversion ($T$) symmetries is always $S \, P \, O \, T = - 1$.~\cite{AMBS_Multiband2013,Triola2016PRB} This enables, e.g., an odd-$\omega$ spin-singlet $s$-wave superconducting pair amplitude to exist, as long as it is also odd in the orbital index. The same odd-$\omega$ mechanism has also been found in double quantum dots, Rashba wires, and layer systems proximity coupled to a conventional superconductor, where the orbital index is then replaced by the dot, wire, or layer label.~\cite{Sothmann2014, Ebisu2016, Parhizgar2014}
Quite generally, this odd-$\omega$ pairing exists if finite inter-orbital pairing is present, which is pairing where the two electrons in the Cooper pair originate from different orbitals.~\cite{AMBS_Multiband2013,Komendova2015} 
However, one big hurdle for promoting this odd-$\omega$ state is to find superconducting systems with large inter-orbital pairing.

A potentially very exciting prospect for odd-$\omega$ superconductivity is the metallic edge states of two-dimensional topological insulators (TIs), or quantum spin Hall insulators (QSHIs), in close proximity with a conventional superconductor. Such QSHI superconducting hybrid systems have already generated a significant amount of interest due to their promise of harboring Majorana bound states.~\cite{FuKane2008, FuKane2009, AMBS2011QSHI_SC, DK2016Si_SFS} Experimental progress has also been significant, finding signatures of both topological superconductivity and gapless Andreev bound states.~\cite{Veldhorst2012,Hart2014,Pribiag2015,Bocquillon2017}
In terms of odd-$\omega$ superconductivity, the very special helical dispersion of the metallic edge states in a topological insulator has recently been shown to generate an odd-$\omega$ spin-triplet $s$-wave state whenever there is an in-surface gradient of the superconducting order parameter, for example due to a junction or an applied supercurrent.~\cite{AMBS_TI2012}
Moreover, even the simplest models of QSHIs require at least two orbitals, which are always very strongly coupled. This could also open for the possibility of additional odd-$\omega$  superconductivity through interorbital pairing. 
Thus superconducting QSHIs are very attractive systems for discovering multiple co-existing, but different, odd-$\omega$ superconducting states and studying their interplay.

In this paper we investigate all different superconducting states in a QSHI proximity coupled to a conventional spin-singlet $s$-wave superconductor. Even limiting the discussion to the disorder robust $s$-wave spatial symmetry, we very generally find multiple odd-$\omega$ superconducting states present and co-existing. Notably, these states exists without any ferromagnetic regions present, which is very different from the usual situation of odd-$\omega$ $s$-wave superconductivity only appearing in superconductor-ferromagnet junctions or other systems with broken time-reversal symmetry.
More specifically, we use the Kane-Mele model~\cite{KaneMeleQSHI} on a honeycomb lattice as a realization of a QSHI. This model is not only theoretically tractable, but also experimentally realized in silicene,~\cite{Houssa2015Si} the silicon analog of graphene with an experimentally achievable band gap in the bulk, as well as materials based on the heavier elements in group IV, such as germanene and stanene.~\cite{Liu2011SiGe, Liu2011GeSn, Davila2014Ge, Zhu2015Sn, Wang2017Sn}

In superconducting QSHI systems we find, without the need of fine tuning, co-existing odd-$\omega$ spin-triplet (OT) and spin-singlet (OS) $s$-wave superconducting states. The OS pair amplitude is most universal and appears both in the bulk of the heavily doped QSHI and at the edge of finite-sized QSHI ribbons. This state is prominent thanks to a naturally strong inter-orbital pairing in the QSHI. In the bulk a finite electric field is needed in order to provide a necessary asymmetry for producing odd-$\omega$ pairing, but in ribbons this asymmetry is automatically generated by the decay of the edge state into the bulk. In all cases, the OS state is closely following the parameter behavior of the proximity-induced even-$\omega$ spin-singlet (ES) state.
The spin-triplet OT state we find requires a finite in-surface gradient to exist. Even if we do not impose any external gradients, we surprisingly discover that natural atomic scale variations on the armchair (AC) edge provide sufficient in-surface gradients, and thus an OT state exist at the AC edge. However, for the zigzag (ZZ) edge there is only canceling gradients and no OT state. Notable, since the OT state cannot appear in the bulk and only appears on the AC edge, it is fundamentally if very different origin compared to the OS state.
Moreover, the strong inter-orbital pairing also generates an even-$\omega$ spin-triplet (ET) state with $s$-wave symmetry on the AC edge, which is odd in orbital index. In the same way that the OS state tracks the behavior of the ES state, this ET state tracks the behavior of the OT state. Thus we find for AC edge QSHI ribbons pair amplitudes with all possible symmetry combinations present.
Finally, we show that all odd-$\omega$ pair amplitudes in this system preserve time-reversal symmetry. This is due to the time-reversal operator not only changing the sign of time, but also flipping the spin and involving a complex conjugation.

The rest of this paper is organized as follows. In the next section we present the model and the methods used to extract the different pair amplitudes and their properties, including the behavior under time-reversal symmetry. In Sec.~\ref{sec:ResultsBulk} we derive all pair amplitudes present in heavily doped QSHIs where the bulk is also metallic. Then in Sec.~\ref{sec:QSHI} we turn to the topological phase with an insulating bulk and only the edge superconducting and here investigate ribbons with both AC and ZZ edges. Finally in Sec.~\ref{sec:Conclusions} we summarize our results.

% ------------------------------------ %
% MODEL:
% ------------------------------------ %
\section{\label{sec:Model}Model}

% Lattice model
\subsection{\label{subsec:ModelHamLatt}Lattice Hamiltonian}
Having silicene and its close relatives in mind, we model a QSHI on a honeycomb lattice using the Kane-Mele model.~\cite{KaneMeleQSHI,Drummond2012, Ezawa_NJP2012} The normal part of the Hamiltonian describing the band structure is given by
\begin{eqnarray}
\label{eq:HamNorm}
\mathcal{H}_{0} & = & t \, \sum_{\langle i, j\rangle\alpha} c^{\dagger}_{i\alpha} \, c_{j\alpha} + \frac{\mathrm{i} \, \kmpar}{3 \, \sqrt{3}} \, \sum_{\langle\langle i, j \rangle\rangle\alpha\beta} \nu_{i j} c^{\dagger}_{i\alpha} \left(\sigma_{z}\right)_{\alpha\beta} c_{j\beta} \nonumber \\
 & - & \sum_{i\alpha} \tilde{\mu}_{i} c^{\dagger}_{i\alpha} c_{i\alpha},
\end{eqnarray}
where $c^{\dagger}_{i\alpha}$ is the creation operator at site $i$ with spin-$z$ component $\alpha$. Here, $t$ is the nearest-neighbor (NN) hopping parameter and $\kmpar$ is the spin-orbit coupling present on next-NNs, with $\nu_{i j} = \pm 1$ depending on whether the turn from site $j$ to site $i$ is in the clockwise or counterclockwise direction. 
Notably, the two non-equivalent atoms per unit cell of the honeycomb lattice provide a multi-orbital basis even when considering only this minimal model of one ($p_z$) orbital per atom. Thus the NN hopping $t$ dominates the kinetic energy, while at the same time providing an extremely strongly inter-orbital coupling. 
The additional spin-orbit coupling opens a non-trivial bulk gap near the $K$ and $K'$ points of the first Brillouin zone, causing the system to enter a QSHI phase, which hosts helical and spin-polarized edge states. We also allow for a sublattice-dependent chemical potential $\tilde{\mu}_{i} = \mu + \zeta_{i} \, \stpar$, with $\zeta_{i} = \pm 1$ depending on whether the site belongs to sublattice $A$ or $B$. The sublattice staggering $\stpar$ is directly proportional to an electric field applied perpendicular to the honeycomb sheet and is thus a highly tunable parameter, a fact we will use extensively in this paper. This tunable sublattice asymmetry is present due to a finite buckling of the honeycomb lattice for elements heavier than carbon.
For strong staggering $\stpar$, the bulk energy gap is notably decreased around $K$ for spin-up electrons and around $K'$ for spin-down electrons. At $\stpar = \kmpar$ the bulk gap eventually closes, causing a topological phase transition into a trivial insulating regime.
We will both focus on extensive sheets, with translational invariance in both directions, and ribbons that have translational invariance only along one spatial direction. These ribbons are characterized by their edge orientation and we consider both ZZ and AC edges, which are the two most common terminations of the honeycomb lattice.

\begin{figure}[htb]
\includegraphics[scale=0.65]{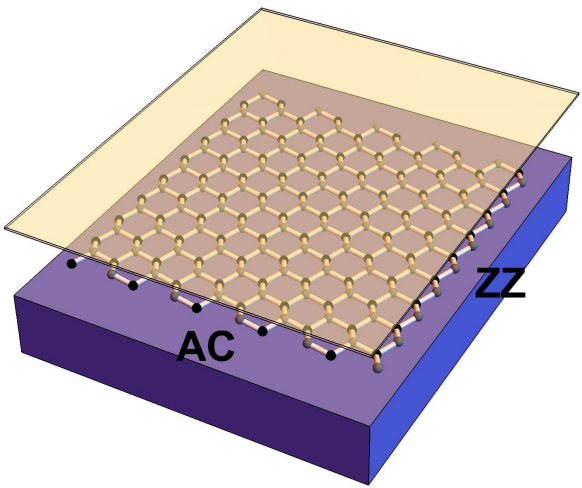}
\caption{\label{fig:Diagram}
Schematic picture of the system with a conventional spin-singlet $s$-wave superconductor substrate (blue) and a honeycomb QSHI sheet on top with $A$ site atoms (black) and $B$ sites (gray), as well as ZZ or AC edges indicated. Top semi-transparent electrode is a gate controlling $\stpar$.}
\end{figure}

The QSHI in Eq.~\eqref{eq:HamNorm} is put in close proximity to a conventional spin-singlet $s$-wave superconductor. 
The general geometry is sketched in Fig.~\ref{fig:Diagram}, where the QSHI is grown on top of a conventional superconductor.
%
% FIGURE:
To model the anomalous self-energy induced in the QSHI from the superconducting gate, we use a constant attractive Hubbard-$U$ interaction which captures the induced attraction:
\begin{equation}
\label{eq:Hubbard}
\mathcal{H}_{\rm{Hub}} = -U \, \sum_{i} c^{\dagger}_{i\uparrow} \, c^{\dagger}_{i\downarrow} \, c_{i\downarrow} \, c_{i\uparrow}.
\end{equation}
Treating this term as usual within mean-field theory, there is then an effective site-dependent superconducting order parameter:
\begin{equation}
\label{eq:Delta}
\Delta_{i} = -U \, \left\langle c_{i\downarrow} \, c_{i\uparrow} \right\rangle
\end{equation}
on every site such that the effective superconducting Hamiltonian part for the QSHI becomes
\begin{equation}
\label{eq:HamSC}
\mathcal{H}_{\rm{SC}} = \sum_{i} \Delta_{i} c^{\dagger}_{i\uparrow} \, c^{\dagger}_{i\downarrow} + \rm{H.c.}.
\end{equation} 
Although the induced pairing interaction $U$ is set constant throughout the system due to a uniform superconducting contact, it is important to notice that $\Delta_{i}$ can vary strongly in space, especially between the edge and bulk regions at low doping, as these are then metallic and insulating, respectively. Also the sublattice staggering can, as we will see, induce a strong sublattice asymmetry in the superconducting order parameter. It is therefore important to solve self-consistently for $\Delta_i$, and not just assume a constant order parameter throughout the system. We do this by first diagonalizing the large matrix $\mathcal{H} = \mathcal{H}_{0} + \mathcal{H}_{\rm{SC}}$ and then evaluating a new $\Delta_i$ using the resulting eigenvectors and eigenvalues. The newly calculated $\Delta_i$ is then used in $\mathcal{H}$ and the process is reiterated until the maximal difference at every site in $\Delta_i$ between successive iterations is below a certain pre-determined convergence criterion (here set to $0.5 \times 10^{-3} \, t$).

% pair amplitude:
\subsection{\label{subsec:ModelOddwF}Superconducting pair amplitudes}
We are in this paper primarily focused on the different superconducting states or pair amplitudes induced into the QSHI. While the superconductor itself only hosts a conventional spin-singlet $s$-wave superconductivity, we will in this paper show that the highly non-trivial band structure of the QSHI results in a plethora of different proximity-induced pair amplitudes. These are all given by the real time-ordered anomalous Green's function in the QSHI
\begin{equation}
\label{eq:Flattice}
\mathrm{i} \,F_{\alpha\beta}(i , j, t) = \left\langle T_{t} c_{i\alpha}(t) \, c_{j\beta}(0) \right\rangle,
\end{equation}
Due to the Fermi-Dirac statistics of the Cooper pairs, the anomalous Green's function needs to satisfy the full antisymmetry condition $F_{\alpha\beta}(i , j, t) = - F_{\beta\alpha}(j , i, -t)$.
The antisymmetry condition makes it convenient to analyze $F$ in terms of its time, spin, and spatial symmetry components.
We can define pair amplitudes that are either even ($E$) or odd ($O$) in the time variable, or equivalently frequency ($\omega$), as\cite{AMBS_TI2012, Dahal2009, AVB2}
\begin{align}
\label{eq:FTlattice}
F^{E}_{\alpha\beta}(i, j) & = \lim_{t \rightarrow 0} \frac{1}{2} \, \left[ F_{\alpha\beta}(i, j, t) + F_{\alpha\beta}(i, j , - t) \right],  \\
F^{O}_{\alpha\beta}(i, j) & = \lim_{t \rightarrow 0} \frac{\partial}{\partial t} \, \left\lbrace \frac{1}{2} \, \left[ F_{\alpha\beta}(i, j, t) - F_{\alpha\beta}(i, j, -t) \right] \right\rbrace. \nonumber
\end{align}
With this construction $\mathrm{i}F^{E}$ is just the traditional expression for the even-$\omega$ pair amplitude, e.g., $\langle c_{i\uparrow} c_{i\downarrow}\rangle$ for the spin-singlet $s$-wave state, while the time derivative in the expression for $F^{O}$ guarantees that it is only non-zero for odd time dependence.
Moreover, using a time derivative allows us to still only work with equal-time expectation values even for the odd-$\omega$ response.
Practically we can always calculate \req{eq:FTlattice} by constructing the time-dependent eigenstates of the lattice Hamiltonian $\mathcal{H} = \mathcal{H}_0 + \mathcal{H}_{\rm SC}$, which are easily accessible since we also know all eigenstates and energies.
Furthermore, for any $F$ we construct pair amplitudes that have spin-singlet ($S$) or mixed spin-triplet ($T$) symmetry using
\begin{align}
\label{eq:spindecomp}
F^{S} & =  \frac{1}{2} \left(F_{\uparrow\downarrow} - F_{\downarrow\uparrow} \right),\nonumber \\
F^{T} & =  \frac{1}{2} \left(F_{\uparrow\downarrow} + F_{\downarrow\uparrow} \right).
\end{align}
Note that equal-spin amplitudes are not present unless we apply a magnetic field. 
Finally, for the spatial symmetries we concentrate on $s$-wave symmetry, since higher angular momentum states are much less stable in the presence of disorder. We extract both the on-site $s$-wave and extended $s$-wave ($s^+$) pair amplitudes according to 
\begin{align}
\label{eq:swave}
F_{s}(i) & = F(i , i), \nonumber \\
F_{s^{+}}(i) & = \sum_{j \in \langle i, j \rangle} F(i, j).
\end{align}
Note that the $s^+$-wave state is still associated with a site, although it resides on the bonds emanating out from that site. We here limit ourselves to NN bonds, but longer-range spatial correlations show the same behavior.

% CHECK THIS SECTION AGAIN:
% TIME-REVERSAL:
\subsection{\label{subsec:TRS}Time-reversal symmetry}
When considering pair amplitudes with different parity with respect to the time coordinate, it is natural to ask the question of how such states behave under time-reversal symmetry. Here, we derive the necessary condition a general pair amplitude must satisfy in order to respect time-reversal symmetry, provided that the many-body Hamiltonian is time-reversal invariant.

For any, not necessarily Hermitian operator $A$ and a many-body Hamiltonian that is invariant under time reversal, the action of the time-reversal operator on its expectation value is given by
\[
\begin{array}{rcl}
\mathcal{T} \, \mathrm{Tr} \left\lbrace \rho \, A \right\rbrace \, \mathcal{T}^{-1} & = & \sum_{n} \mathrm{e}^{\beta \, \left(\Omega - E_{n} \right)} \, \mathcal{T} \, \left\langle E_{n} \right \vert A \left \vert E_{n} \right\rangle \, \mathcal{T}^{-1} \\
 & = & \sum_{n} \mathrm{e}^{\beta \, \left(\Omega - \tilde{E}_{n} \right)} \, \left\langle \tilde{E}_{n} \right \vert \mathcal{T} \, A^{\dagger} \, \mathcal{T}^{-1} \left \vert \tilde{E}_{n} \right\rangle \\
 & = & \mathrm{Tr} \left\lbrace \rho \, \mathcal{T} \, A^{\dagger} \, \mathcal{T}^{-1} \right\rbrace.
\end{array}
\]
Here, in the second line we used the fact that the eigenvalues $E_{n}$ of the many-body Hamiltonian are invariant under time reversal, $E_{n} = \tilde{E}_{n}$, if the Hamiltonian itself respects this symmetry and, with them, the free energy (grand thermodynamic potential) $\Omega$. In the above, we had used the canonical (grand canonical) distribution at finite temperature, but the argument holds even in the zero-temperature limit, provided that the ground state of the system, together with the Hamiltonian, is invariant under time reversal.
Then, having in mind \req{eq:Flattice}, consider the operator
\[
A = T_{t} \left[ c_{i\alpha}(t) \, c_{j\beta}(0) \right],
\]
which is not Hermitian. After some algebraic manipulations and using the action of the time-reversal operator on spin-$1/2$ operators, we obtain
\[
\mathcal{T} \, A^{\dagger} \, \mathcal{T}^{-1} = (\mathrm{i} \sigma_{2})_{\alpha\gamma} \, (\mathrm{i} \sigma_{2})_{\beta\delta} \, \left[ \mathrm{T}_{t} \, \left( c_{i\gamma}(-t) \, c_{j\delta}(0) \right) \right]^{\dagger}.
\]
Finally, using the identity $\mathrm{Tr} \left\lbrace \rho \, X^{\dagger} \right\rbrace = \left(\mathrm{Tr} \left\lbrace \rho \, X \right\rbrace\right)^{\ast}$, we may combine the above steps on the definition \req{eq:Flattice} to obtain:
\begin{equation}
\label{eq:FLatticeTRS}
\mathcal{T} \, \mathrm{i} \, F_{\alpha\beta}(i, j, t) \, \mathcal{T}^{-1} = \left(\mathrm{i} \sigma_{2}\right)_{\alpha\gamma} \, \left(\mathrm{i} \sigma_{2}\right)_{\beta\delta} \left[\mathrm{i} \, F_{\gamma\delta}(i, j, -t)\right]^{\ast}.
\end{equation}
\req{eq:FLatticeTRS} forms a constraint all general pair amplitudes have to satisfy for a system to obey time-reversal symmetry. We note that, besides changing the sign of the time coordinate, the action of the time-reversal operator also involves acting on the spin indices, as well as taking the complex conjugate of the expectation value, as it is an anti-linear operator. However, it does not change the spatial coordinates, as is to be expected.

Using the separations into even and odd $\omega$ in \req{eq:FTlattice} and the spin decomposition in \req{eq:spindecomp}, time-reversal invariance implies the following relations
\begin{align}
\label{eq:TRSconds}
\left[ \mathrm{i} \, F^{ES}(i, j) \right]^{\ast} & =  \mathrm{i} \, F^{ES}(i, j), \nonumber \\
\left[ \mathrm{i} \, F^{ET}(i, j) \right]^{\ast} & =  -\mathrm{i} \, F^{ET}(i, j), \nonumber\\
\left[ \mathrm{i} \, F^{OS}(i, j) \right]^{\ast} & = - \mathrm{i} \, F^{OS}(i, j), \nonumber\\
\left[ \mathrm{i} \, F^{OT}(i, j) \right]^{\ast} & =  \mathrm{i} \, F^{OT}(i, j). 
\end{align}
As will be explicitly demonstrated in the following sections, the conditions \reqs{eq:TRSconds} are indeed always fulfilled despite multiple different odd-$\omega$ pair amplitudes often being present. Thus, a QSHI proximity coupled to a conventional superconductor always preserves time-reversal symmetry, even though odd-$\omega$ superconductivity is present.

%
% BULK:
\subsection{\label{subsec:ModelKspace}Translationally invariant bulk}
The formalism developed so far is suitable to analyze everything from large sheets to thin nanoribbons. However, in the case of a very large sheet, there is preserved translational symmetry, and transforming to Fourier space diagonalizes the Hamiltonian. 
We use as real-space lattice vectors $\bm{a}_{1/2} = \pm\frac{3 a}{2} \, \hat{\bm{x}} + \frac{a \sqrt{3}}{2} \, \hat{\bm{y}}$, which gives the reciprocal-lattice vectors $\bm{b}_{1/2} = \pm \frac{1}{3 a} \, \hat{\bm{x}} + \frac{1}{a \sqrt{3}} \, \hat{\bm{y}}$, where $a$ is the NN distance. Then, the reciprocal $k$ vector is parametrized as $\bm{k} = k_{1} \, \bm{b}_{1} + k_{2} \, \bm{b}_{2}$, with $k_{1/2} \in [-\pi, \pi]$.
Due to the special form of the spin-orbit coupling in \req{eq:HamNorm}, $S_z$ is a good quantum number and the full Hamiltonian therefore decomposes into separate two spin blocks. Using the Nambu spinor $\Psi_{\bm{k}\sigma}^\top = \left( c_{A,\bm{k},\sigma}, c_{B,\bm{k},\sigma},c^{\dagger}_{A,-\bm{k},-\sigma}, c^{\dagger}_{B,-\bm{k},-\sigma} \right)$, where $A$ and $B$ indicate the two sublattice sites, the full Hamiltonian has the block structure
\begin{align}
\mathcal{H}_{\rm bulk}  = & \frac{1}{2}\sum_{\bm{k}\sigma} \Psi^{\dagger}_{\bm{k}\sigma} \cdot \check{H}_{\sigma}(\bm{k}) \cdot \Psi_{\bm{k}\sigma}, \nonumber \\
\check{H}_{\sigma}(\bm{k})  = & \hat{\tau}_{3} \otimes \left(\bm{p}_{\sigma}(\bm{k}) \cdot \hat{\bm{\rho}} \right) - \mu \, \hat{\tau}_{3} \otimes \hat{\rho}_{0} \nonumber \\
 & +  \sigma \, \Delta_{+} \, \hat{\tau}_{1} \otimes \hat{\rho}_{0} + \sigma \, \Delta_{-} \, \hat{\tau}_{1} \otimes \hat{\rho}_{3},
\end{align}
where $\hat{\tau}_{\mu}$ and $\hat{\rho}_{\nu}$ are Pauli matrices in the Nambu and sublattice, or equivalently orbital, subspaces, respectively. We have here defined
\begin{equation}
\label{eq:Dvector}
\bm{p}_{\sigma}(\bm{k}) = \left( \begin{array}{c@{,}c@{,}c}
t \, \Re (\gamma_{\bm{k}} ) & -t \, \Im (\gamma_{\bm{k}}) & \sigma \kmpar f_{\bm{k}} - \stpar
\end{array} \right),
\end{equation}
where the NN and next-NN modulation factors for the honeycomb lattice are given  by
\begin{align}
\gamma_{\bm{k}} & =  1 + e^{\mathrm{i} \, k_{1}} + e^{-\mathrm{i} \, k_{2}}, \nonumber \\
f_{\bm{k}} & =  \frac{2}{3 \, \sqrt{3}} \, \left[ \sin(k_1) + \sin(k_2) - \sin(k_{1} + k_{2}) \right],
\end{align}
respectively.
Also, it is convenient to work with the symmetric and anti-symmetric sublattice combinations of the order parameter and we therefore also define
\begin{equation}
\label{eq:Deltas}
\Delta_{\pm} = \frac{\Delta_{A} \pm \Delta_{B}}{2}.
\end{equation}
Here $\Delta_{A(B)}$ is the order parameter in sublattice $A$ ($B$). 

% ------------------------------------ %
% METALLIC BULK
% ------------------------------------ %
\section{\label{sec:ResultsBulk}Metallic bulk}
Having developed all the necessary formalism in the previous section, we start by analyzing the superconducting state in the heavily doped regime of the QSHI. At heavy doping, the Fermi level is firmly situated within the conduction (or valence) band of the QSHI and the whole material is a metal. The QSHI edge states are thus not important for the low-energy physics and it is sufficient to analyze the bulk system. In this limit there should be no spatial variations of the proximity-induced superconducting order parameter between unit cells, such that the system has fully translational invariance. Note, however, that a finite staggering $\stpar$ creates a sublattice difference within each unit cell and thus we are required to treat $\Delta_A$ and $\Delta_B$ independently.

% ANALYTICAL RESULTS:
\subsection{\label{subsec:BulkA}Analytical results}
In case of full translational invariance, the anomalous Green's function of the system may be evaluated analytically using the formalism developed in Sec.~\ref{subsec:ModelKspace}. This offers a detailed picture of both the $\bm{k}$ dependence and the parameter dependence of each of the induced pair amplitudes. 
The anomalous Green's function $\hat{F}_{\bm{k}, \sigma}(z)$ is just the particle-hole submatrix of the full Green's-function matrix $\check{G}_{\sigma}(\bm{k}, z) = \left(z \, \check{1} - \check{H}_{\sigma}(\bm{k}) \right)^{-1}$ written in matrix form as
\begin{equation}
\label{eq:fullG}
\check{G}_{\sigma}(\bm{k}, z) = \left(\begin{array}{c|c}
\hat{G}_{p, \sigma}(\bm{k}, z) & \hat{F}_{\sigma}(\bm{k}, z) \\
\hline
\hat{\bar{F}}(\bm{k}, z) & \hat{G}_{h, \sigma}(\bm{k}, z)
\end{array} \right),
\end{equation}
which has the orbital (i.e., sublattice) structure
\[
\hat{F}_{\sigma}(\bm{k}, z) = \frac{1}{D_{\sigma}(\bm{k}, z)} \, \left(\begin{array}{cc}
 N_{\sigma, AA}(\bm{k}, z) & N_{\sigma, AB}(\bm{k}, z) \\
 N_{\sigma, BA}(\bm{k}, z) & N_{\sigma, BB}(\bm{k}, z)
\end{array}\right).
\]
Explicitly solving for the anomalous Green's function we find for the numerators
\begin{widetext}
\begin{align}
\label{eq:Fnumerators}
N_{\sigma, AA}(\bm{k}, z) & =  \sigma \, \left(\Delta_{+} + \Delta_{-} \right) \, \left[ z^2 - \left(\sigma \, \kmpar \, f_{\bm{k}} - \stpar + \mu \right)^2 - \left(\Delta_{+} - \Delta_{-}\right)^2 \right] - \sigma \, \left( \Delta_{+} - \Delta_{-} \right) \, t^2 \, \vert \gamma_{\bm{k}} \vert^{2}, \nonumber \\
N_{\sigma, AB}(\bm{k}, z) & =  - 2\sigma \, t \, \gamma_{\bm{k}} \, \left[ \left(\sigma \, \kmpar f_{\bm{k}} - \stpar + z \right) \, \Delta_{-} + \mu \, \Delta_{+} \right], \nonumber \\
N_{\sigma, BA}(\bm{k}, z) & =  -2 \sigma \, t \gamma^{\ast}_{\bm{k}} \, \left[ \left( \sigma \, \kmpar \, f_{\bm{k}} - \stpar - z \right) \, \Delta_{-}  + \mu \, \Delta_{+} \right], \nonumber \\
N_{\sigma, BB}(\bm{k}, z) & = \sigma \, \left(\Delta_{+} - \Delta_{-} \right) \, \left[ z^2 - \left(\sigma \, \kmpar \, f_{\bm{k}} - \stpar - \mu \right)^2 - \left(\Delta_{+} + \Delta_{-}\right)^2 \right] - \sigma \, \left( \Delta_{+} + \Delta_{-} \right) \, t^2 \, \vert \gamma_{\bm{k}} \vert^{2},
\end{align}
while the denominator is a biquadratic function of $z$:
\begin{align}
\label{eq:GreenDenom}
D_{\sigma}(\bm{k}, z) \! =  \! \left[ z^{2} \!- \! t^2 \, \vert \gamma_{\bm{k}} \vert^{2} \!- \! \left(\sigma \kmpar f_{\bm{k}} - \stpar \right)^2 \!- \! \mu^2 \!- \! \Delta^{2}_{+} \!- \! \Delta^{2}_{-} \right]^2  \!- \! 4 \left[ t^2 \, \vert \gamma_{\bm{k}} \vert^{2} \! \left( \mu^2 + \Delta^{2}_{-} \right) + \left( \left(\sigma \, \kmpar f_{\bm{k}} - \stpar \right) \! \mu - \Delta_{+}  \Delta_{-} \right)^2 \right]\!.
\end{align}
It is clear that the only terms in \reqs{eq:Fnumerators} that are odd in $z$ (and thus in frequency) are the two inter-orbital amplitudes $N_{\sigma,AB}$ and $N_{\sigma,BA}$. Thus any odd-$\omega$ pair amplitudes must be generated by these terms, since the denominator is an even function of $z$.
Focusing on these inter-orbital terms and constructing the odd-$\omega$ spin-singlet (OS superscript) and spin-triplet (OT superscript) pair amplitudes, the only finite terms are those that are odd in orbital index and they are given by 
\begin{eqnarray}
\label{eq:oddFsinglet}
F^{OS}(\bm{k}, z) & = & \frac{1}{2} \, \left[F^{O}_{\uparrow,AB}(\bm{k}, z) - F^{O}_{\downarrow,AB}(\bm{k}, z) \right] = - z\frac{2 \, t \, \Delta_{-}  \, \gamma_{\bm{k}}}{D_{\uparrow}(\bm{k}, z) \, D_{\downarrow}(\bm{k}, z)} \, \left\lbrace \left(z^{2} - \kmpar^{2} \, f^{2}_{\bm{k}} - \stpar^{2} - \mu^{2} - \Delta^{2}_{+} - \Delta^{2}_{-} \right)^{2} \right. \nonumber \\
 & - & \left. 4 \, \left[ \kmpar^{2} \, f^{2}_{\bm{k}} \left( \mu^{2} - \stpar^{2} \right) + t^{2} \, \vert \gamma_{\bm{k}} \vert^{2} \, \left( \mu^{2} + \Delta^{2}_{-} \right) + \left(2 \mu \, \stpar + \Delta_{+} \, \Delta_{-} \right)^{2} \right]\right\rbrace,
\end{eqnarray}
\begin{eqnarray}
\label{eq:oddFtriplet}
F^{OT}(\bm{k}, z) & = & \frac{1}{2} \, \left[F^{O}_{\uparrow,AB}(\bm{k}, z) + F^{O}_{\downarrow,AB}(\bm{k}, z) \right] \nonumber \\
 & = & -z\frac{8 t \, \kmpar \, \Delta_{-} \, \gamma_{\bm{k}} \, f_{\bm{k}}}{D_{\uparrow}(\bm{k}, z) \, D_{\downarrow}(\bm{k}, z)} \, \left[ \stpar \, \left( z^{2} - t^{2} \, \vert \gamma_{\bm{k}} \vert^{2} - \kmpar^{2} \, f^{2}_{\bm{k}} - \stpar^{2} + \mu^{2} - \Delta^{2}_{+} - \Delta^{2}_{-} \right) + 2 \, \mu \, \Delta_{+} \, \Delta_{-} \right].
\end{eqnarray}
\end{widetext}
The odd-$\omega$ spin-singlet pair amplitude in \req{eq:oddFsinglet} is an even function of $\bm{k}$ with predominantly $s^{+}$-wave symmetry. This is due to the NN prefactor $\gamma_{\bm{k}}$. This is fully consistent with Fermi-Dirac statistics, with the oddness in orbital space canceling the oddness in frequency, while keeping the spin-singlet $s$-wave symmetry from the external superconductor. 
This odd-$\omega$ state is generated by a finite interorbital hybridization $t$ and an order-parameter difference between the two sublattices, i.e., a finite $\Delta_{-}$. This is consistent with earlier results for two-band models, where finite inter-band hybridization has been found to generate odd-$\omega$ pairing when an asymmetry is present between the two intraband order parameters~\cite{AMBS_Multiband2013,Komendova2015,Komendova2017}.

For the odd-$\omega$ spin-triplet pair amplitude in \req{eq:oddFtriplet} we instead find an overall odd $\bm{k}$ dependence, which is mainly $p$ wave due to the prefactor $\gamma_{\bm{k}} \, f_{\bm{k}}$. Thus the Fermi-Dirac statistics of the Cooper pair is fulfilled also in this case.
We note that the spin-orbit coupling in the QSHI generates a spin-triplet $p$-wave state also in the even-frequency regime,~\cite{FuKane2009,AMBS2011QSHI_SC,AMBS_TI-SCprox2013} although the orbital symmetry is then even. The OT pair amplitude is also directly proportional to the spin-orbit coupling $\kmpar$, in addition to the same dependence on $t$ and $\Delta_-$ as for the OS component. Based on this, we conclude that the odd-$\omega$ states in Eqs.~\eqref{eq:oddFsinglet} and \eqref{eq:oddFtriplet} are the odd-$\omega$ odd inter-orbital companions to their even-frequency even-orbital counterparts. This shows explicitly that the reciprocity between parity in time and orbital space is independent of the spin and spatial parity. For silicene and related QSHIs, both of these odd-$\omega$ pair amplitudes are present in the metallic state as soon as there is a finite $\Delta_-$.

% NUMERICAL RESULTS
\subsection{\label{subsec:ResultsBulkN}Numerical calculations}
Here we perform complementary self-consistent calculations in the metallic bulk regime to show how a finite $\Delta_-$ is present as soon as there is finite staggering $\stpar$ and how this generates odd-$\omega$ pairing.
 For these calculations we choose a high enough value of $\mu$ such that the Fermi level is above the bulk gap $E \leq \kmpar$, but still below the van Hove singularity at $E = t$, so as to not cause an artificial enhancement of superconductivity.
  
 % FIGURE
\begin{figure}[htb]
\includegraphics[scale=1.0]{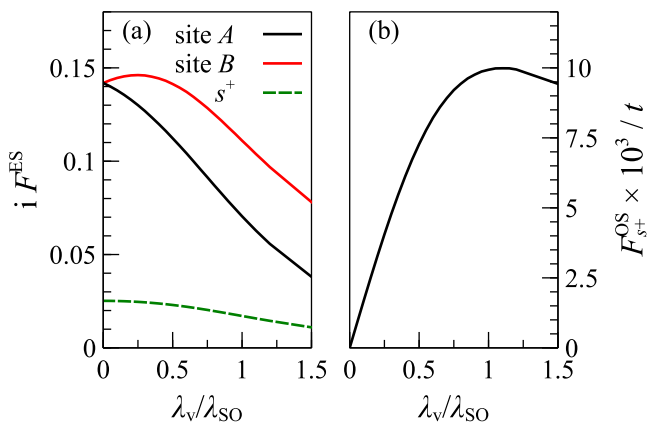}
\caption{\label{fig:BulkPairAmps}
Pair amplitudes for a metallic QSHI as a function of sublattice staggering $\stpar$. Here the parameters are $\kmpar/t = 0.50$, $U/t = 2.0$, and $\mu/t = 0.70$. (a): ES on-site $s$-wave and NN $s^{+}$-wave, (b): OS NN $s^{+}$-wave.}
\end{figure}
In Fig.~\ref{fig:BulkPairAmps}, we plot all the non-zero $s$-wave pair amplitudes. Figure~\ref{fig:BulkPairAmps}(a) depicts how the ES on-site $s$-wave pair amplitudes on the $A$ and $B$ sites vary as the staggering $\stpar$ increases. At zero staggering they are equal and thus $\Delta_A = \Delta_B$ through \req{eq:Delta}. However, they become unequal for finite staggering. This is expected, since finite staggering results in different local density of states (LDOS) around the Fermi level for each sublattice. The resulting different pair amplitudes give $\Delta_A\neq \Delta_B$, which is the  criterion analytically derived above for generating odd-$\omega$ pairing. In Fig~\ref{fig:BulkPairAmps}(b) we then plot the only non-zero odd-$\omega$ pair amplitude, which has spin-singlet extended $s^+$ symmetry. The $s^+$ symmetry allows for an orbital dependence and we have confirmed that it is odd in the orbital index, thus fully obeying Fermi-Dirac statistics. This OS pair amplitude tracks $\Delta_-$ and is therefore the pair amplitude found in \req{eq:oddFsinglet}. Note that $F^{OS}$ is fully real and thus it does not break time-reversal symmetry according to \req{eq:TRSconds}, despite the odd-$\omega$ dependence.
There is also a finite even-$\omega$ spin-singlet $s^+$-wave component at all staggerings. This pair amplitude can exist because it belongs to the same irreducible representation (the identity) of the point group as the $s$-wave pair amplitude.
We also note that nothing special happens at the critical staggering $\stpar = \kmpar$ for the QSHI topological phase, since the system is already in the metallic regime.

% ------------------------------------ %
% QSHI RIBBONS
% ------------------------------------ %
\section{\label{sec:QSHI}QSHI ribbons}
We now focus on the low doping regime where the Fermi levels fall within the bulk energy gap. The only low-energy excitations are then those of the topologically protected edge states.
We here set the chemical potential slightly away from zero to break the (accidental) particle-hole symmetry of the normal state, but still well within the bulk energy gap. We assume, very realistically, that the proximity-induced superconductivity is too weak to produce a superconducting gap in the insulating bulk interior, and we thus use a $U$ such that superconductivity is only present in the edge states, despite $U$ being finite and constant throughout the ribbon. 
In order to model the edge states we consider semi-infinite ribbons of the QSHI with translational symmetry in the direction along the ribbon and wide enough, such that the two edges do not hybridize.

For easy comparison, we use the same $U/t =2$ in the ribbon as in the metallic bulk case studied above. By simply changing the chemical potential between these two cases we achieve edge-only superconductivity in the ribbon, but bulk superconductivity in the metallic case. 
We note that this choice of $U$ produces a somewhat large $\Delta$, which is required due to the computational limitations of the size of the systems we can study. In short, the system size needs to be at least a few times the superconducting coherence length $\xi=v_F/\Delta$ for a self-consistent approach to be valid. Despite this limitation with self-consistent calculations, multiple previous studies in related systems have still generated experimentally reliable results (see e.g., Refs.~\onlinecite{AMBS_JL2010, VanHarlingen2016}).
Moreover, all our results are qualitative, do not change with the precise value of $\Delta$, and are also in agreement with analytical work in the metallic bulk case. Thus, the precise choice $U/t$ is not important for our results and conclusions.

% ZIGZAG EDGES:
\subsection{\label{subsec:ResultsZigZag}Ribbons with zigzag edges}
\begin{figure}[htb]
\includegraphics[width=\linewidth]{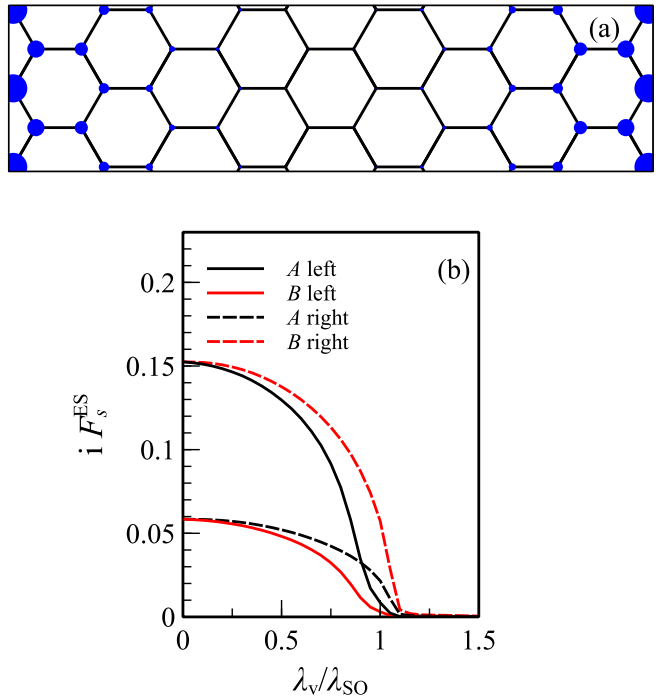}
\caption{\label{fig:ZZESS} ES on-site $s$-wave pair amplitude for a ZZ ribbon. We use the same parameters as in the metallic bulk case, except the chemical potential is $\mu/t = 0.05$. (a): Spatial intensity profile of the pair amplitude for the whole ribbon at $\stpar/\kmpar = 0.0$. The radius of each circle is proportional to the magnitude, while positive (negative) sign is coded in blue (red).
(b): Evolution of pair amplitudes with $\stpar$ on edge sites ($A$ left; $B$, right) and one layer deeper ($A$ right; $B$, left).
}
\end{figure}

% ES s-wave:
First we consider ribbons with ZZ edges. In Fig.~\ref{fig:ZZESS} we analyze the conventional ES on-site $s$-wave pair amplitude, which gives superconducting order parameters $\Delta_i$ through \req{eq:Delta}.
% FIGURE
As displayed in the intensity plot in Fig.~\ref{fig:ZZESS}(a), this pair amplitude takes its maximum value on the edge atoms. However, because the left edge truncates with an $A$ site, and the right edge with a $B$ site, the edge atom pair amplitudes evolve slightly differently with $\stpar$, as shown in Fig.~\ref{fig:ZZESS}(b). This is due to the finite chemical potential $\mu$. 
The decay of the pair amplitude into the interior of the ribbon is very rapid. For this particular choice of parameters, the pair amplitude has already decreased $2.5$ times to the next atomic line. Note that the decay length is primarily set by the spatial extent of the edge state and not the superconducting coherence length.
Moreover, increased staggering causes an overall suppression of superconductivity. This is not surprising, since the edge states are lost at the topological phase transition at $\stpar = \kmpar$, causing superconductivity to vanish in the whole system.

\begin{figure}[htb]
\includegraphics[width=\linewidth]{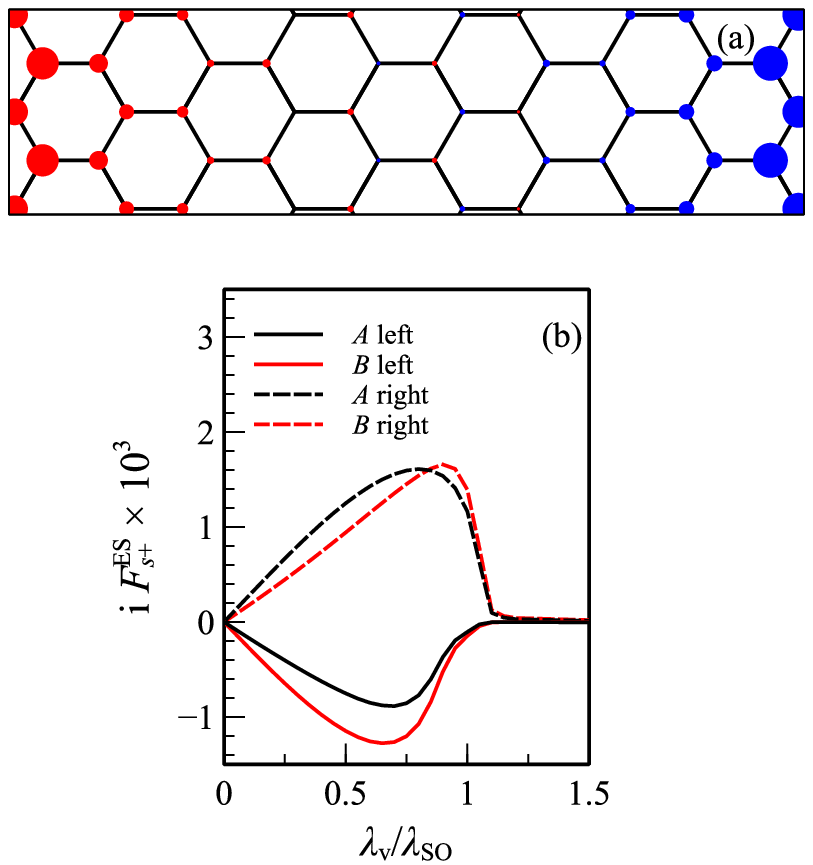}
\caption{\label{fig:ZZESSp} ES NN bond $s^{+}$-wave pair amplitude for the same parameters and geometry as in Fig.~\ref{fig:ZZESS}, apart from $\stpar/\kmpar = 0.65$ in (a).
}
\end{figure}

% ES s+ wave
Just as in the metallic bulk case, the on-site pair amplitude spreads to NN bonds. This sub-dominant ES, $s^{+}$-wave pair amplitude is displayed in Fig.~\ref{fig:ZZESSp}. 
It also decays rapidly into the bulk although it retains quite similar values on the edge atom and its NN site. This is a consequence of $s^+$ pairing being defined on NN bonds and \emph{even} in the exchange in sublattice sites, as required by Fermi-Dirac statistics.
As seen, the sign of the ES $s^{+}$ state is opposite on the two edges. This is possible because the two edges terminate with different sublattice sites. Since switching the sublattice type changes the sign of $\stpar$, we also expect the ES $s^{+}$ state to change sign at $\stpar = 0$, resulting in the zero pair amplitude value at $\stpar = 0$ seen in Fig.~\ref{fig:ZZESSp}(b). Together with vanishing superconductivity at $\stpar = \kmpar$, this explains the dome-shaped curve as a function of staggering.

\begin{figure}[htb]
\includegraphics[width=\linewidth]{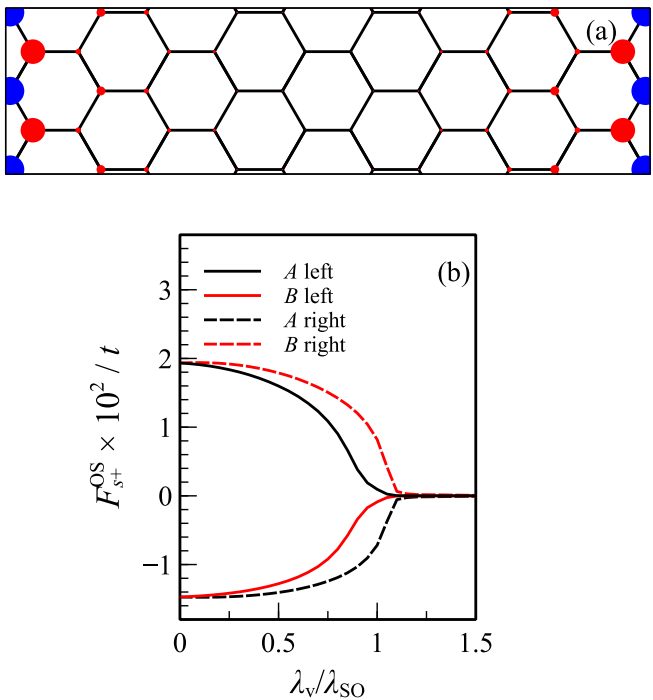}
\caption{\label{fig:ZZOSSp} OS NN bond $s^{+}$-wave pair amplitude for the same parameters and geometry as in Fig.~\ref{fig:ZZESS}.}
\end{figure}

% ODD-W IN ZZ:
Guided by the results from the metallic bulk we also expect odd-$\omega$ pairing in the QSHI ribbons. We have explicitly calculated all possible odd-$\omega$ states with $s$ or $s^+$ symmetry in the ribbon configuration. In Fig.~\ref{fig:ZZOSSp} we show the only non-zero odd-$\omega$ state for a ZZ QSHI ribbon: the OS, $s^+$-wave state, which is odd in orbital index.
Very interestingly, it is finite for all staggering values until superconductivity is lost at the topological phase transition.
This odd-$\omega$ pair amplitude is generated by the same mechanism as the OS state in the metallic bulk and thus requires $\Delta_A \neq \Delta_B$. Still, in a QSHI ribbon it does not vanish at $\stpar =0$; in fact, it is largest for small $\stpar$.
This is possible due to the sublattice sites having different distances to the edge, and thus they have notably different LDOS. This automatically results in a strong asymmetry between $\Delta_A$ and $\Delta_B$, without any need of a finite staggering. In fact, the evolution with $\stpar$ of the OS state tracks that of the ES state. This shows that the OS state on the ZZ edge is only dependent on the very existence of superconductivity at the edge and not on anything else. This lack of a strong dependence on $\stpar$ is also evident from the left and right edges showing the same overall behavior. 
We also note that the oddness under exchange of sublattice sites gives rise to alternating signs of the pair amplitude along the ZZ edge. This is the complete opposite behavior from the even inter-orbital pair amplitude in Fig.~\ref{fig:ZZESSp}.
Similar to the metallic bulk OS component, this pair amplitude is fully real and thus preserves time-reversal symmetry.

% ARMCHAIR:
\subsection{\label{subsec:ResultsArmChair} Ribbons with armchair edges}
The honeycomb lattice is rather unique in that it has two preferential edge directions, ZZ and AC, that have very different geometries. Especially prominent is that the AC edge has atoms of both sublattice sites along the edge. We therefore expect that AC ribbons might behave qualitatively quite different from the ZZ ribbons investigated above.
%
% ES s-wave:
\begin{figure}[htb]
\includegraphics[width=\linewidth]{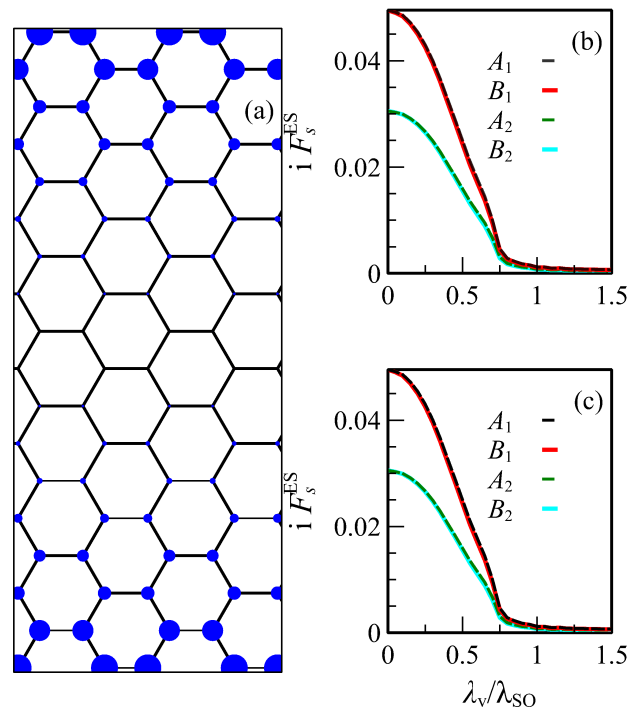}
\caption{\label{fig:ACESS} ES on-site $s$-wave pair amplitude for an AC ribbon for the same parameters as Fig.~\ref{fig:ZZESS}. (a): Spatial intensity profile of the pair amplitude for the whole ribbon at $\stpar/\kmpar = 0.0$. (b, c) Evolution of the pair amplitudes with $\stpar$ on the top (b) and bottom (c) edge on edge sites (index $1$) and one layer beyond (index $2$).
}
\end{figure}
We start investigating AC ribbons by plotting the ES on-site $s$-wave pair amplitude in Fig.~\ref{fig:ACESS}, the equivalent plot to Fig.~\ref{fig:ZZESS} for ZZ ribbons.
For the AC ribbon we find that the top and bottom edges show exactly the same pair amplitude. This is true for all pair amplitudes we have investigated and we will henceforth report results for the bottom edge only.
Moreover, this ES pair amplitude acquires the same value on $A$ and $B$ sites that sit at the same distance from the edge, independent of $\stpar$. This behavior is very different from the ZZ edge and is a consequence of the strong robustness of the  AC edge states against buckling. In fact, the normal state shows no LDOS difference between the two sublattice sites even for finite values of $\stpar$. As a direct consequence, $\Delta_A$ and $\Delta_B$ are the same and only dependent on their distance to the edge. 
Given the same choice of parameters, the magnitude of the ES pair amplitude on the AC edge is about three times lower for the ZZ edge. Also, it only decreases $1.7$ times to the next atomic line, i.e.~slower than for a ZZ edge. These effects are due to the AC QSHI edge states having a lower LDOS, set primarily by a higher edge state Fermi velocity and larger spatial spread into the bulk than the ZZ edge states. 
%
% ES s+ wave:
\begin{figure}[h]
\includegraphics[width=\linewidth]{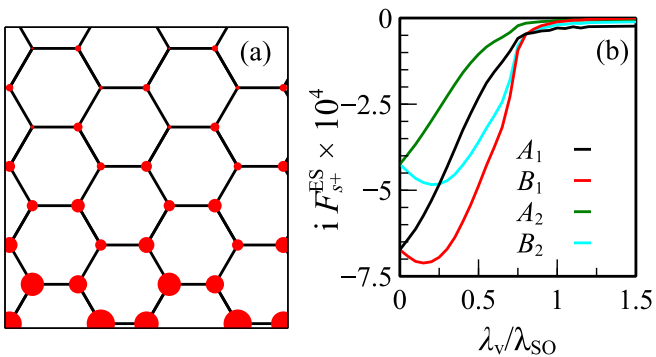}
\caption{\label{fig:ACESSp} ES NN bond $s^{+}$-wave pair amplitude for the same parameters and geometry as in Fig.~\ref{fig:ACESS}, apart from $\stpar/\kmpar = 0.2$ in (a).}
\end{figure}
We also directly investigate the sub-dominant ES, NN bond $s^{+}$-wave pair amplitude in Fig.~\ref{fig:ACESSp}. Here, there is some difference in the $\stpar$ dependence for $A$ and $B$ sublattice sites. This combines with the general decay of the pair amplitude into the bulk to give some more variation in the $s^+$ state with $\stpar$ as compared to the on-site $s$-wave state. 

\begin{figure}[htb]
\includegraphics[width=\linewidth]{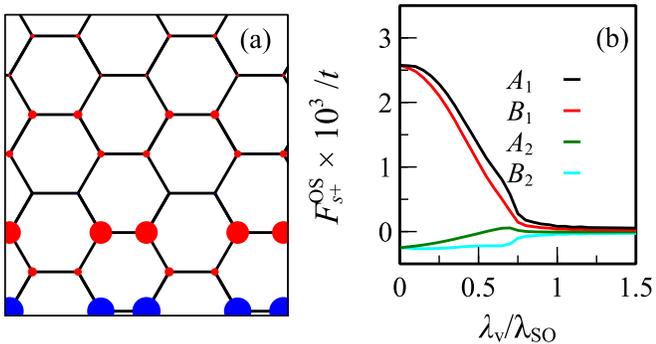}
\caption{\label{fig:ACOSSp} OS NN bond $s^{+}$-wave pair amplitude for the same parameters and geometry as in Fig.~\ref{fig:ACESS}.
}
\end{figure}

 % OS s+ wave:
 The existence of the ES $s^+$-wave state that is necessarily even in orbital index also gives rise to an OS NN bond $s^+$ pair amplitude that is instead odd under orbital exchange. This state is presented in Fig.~\ref{fig:ACOSSp} for the AC edge. 
Both the spatial extent and the evolution with $\stpar$ resemble the behavior of the conventional ES on-site $s$-wave pair amplitude, similarly to the behavior at the ZZ edge. Notably, this OS state does not require a finite staggering to exist. The asymmetry between the $A$ and $B$ sublattice instead comes directly from the spatial decay of the edge state into the bulk. For the AC edge, $A$ and $B$ sites at the same distance from the edge have the same $\Delta$'s, but for the two other NN sites that reside in neighboring layers the sublattice symmetry is broken. We attribute the lower fractional amplitudes of the OS to the ES amplitudes on the AC edge compared to the ZZ edge and also metallic bulk to the fact that not all NN sites show a sublattice asymmetry.

\begin{figure}[htb]
\includegraphics[width=\linewidth]{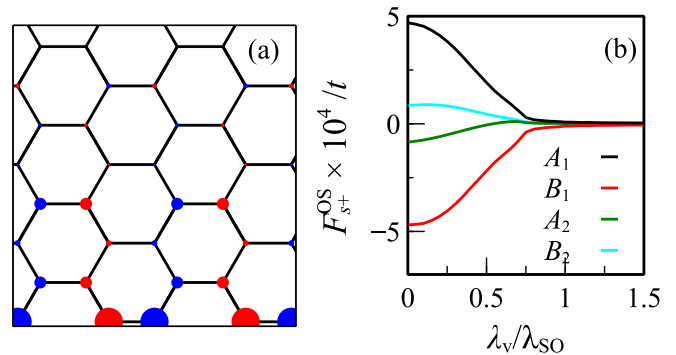}
\caption{\label{fig:ACOTS} OT on-site $s$-wave pair amplitude for the same parameters and geometry as in Fig.~\ref{fig:ACESS}.
}
\end{figure}

\begin{figure}[htb]
\includegraphics[width=\linewidth]{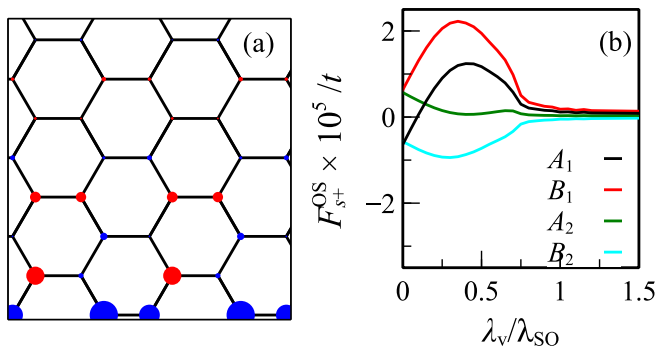}
\caption{\label{fig:ACOTSp} OT NN bond $s^{+}$-wave pair amplitude for the same parameters and geometry as in Fig.~\ref{fig:ACESS}, apart from $\stpar/\kmpar = 0.35$ in (a).}
\end{figure}

% OT ET:
Beyond the OS state, we also find odd-$\omega$ pair amplitudes that have spin-triplet and $s$-wave symmetry at the AC edge, a state {\it not} present at the ZZ edge or in the metallic bulk. It has earlier been shown that OT states with $s$-wave symmetry appear, quite generally, in the surface states of topological insulators whenever there is an in-surface gradient of the conventional (ES on-site $s$-wave) superconducting order.~\cite{AMBS_TI2012}
Here, we demonstrate the presence of such odd-$\omega$ pairing in superconducting silicene and related materials for AC edges. But, importantly, there is here no macroscopic gradient of the superconducting order parameter produced, e.g., by a junction or a supercurrent. Instead, it is simply atomic scale lattice variations that gives rise to the required gradient and thus the OT state.
This can be understood when studying the conventional superconducting order parameters at the ZZ and AC edges, as depicted in Figs.~\ref{fig:ZZESS}(a) and \ref{fig:ACESS}(a), respectively. Both NN atoms to an edge atom on a ZZ edge have the same order-parameter value. Since these two NN atoms are on either side of the edge atom site, the local gradient along the edge for the order parameter is necessarily zero at the edge site. The same argument is true for all sites, even further from the edge, and leads to no OT components on the ZZ edge. This result is independent of the buckling set by $\stpar$.
In contrast, at the AC edge, there is still a zero gradient on the bond sitting at the very edge, but for the other NN site there is a finite projection gradient in the order parameter when projecting along the edge. This gradient is also different for the $A$ and $B$ sites along the edge. Taken together, this results in a finite OT on-site $s$-wave pair amplitude with alternating signs between $A$ and $B$ sites, as clearly shown in Fig.~\ref{fig:ACOTS}.
This state is not very sensitive to the effects of $\stpar$, beyond the overall suppression of superconductivity, also found for the underlying ES state. Similarly to the ES state always existing both on-site and on NN bonds, we also find that the OT state exists both with $s$- and  $s^+$-wave symmetry, as shown in Fig.~\ref{fig:ACOTSp}. Note, however, that the $s^+$ amplitude is suppressed more than an order of magnitude, and thus the OT state has dominantly on-site $s$-wave symmetry.
Thus we conclude that odd-$\omega$ spin-triplet $s$-wave superconductivity appears in superconducting silicene and other related QSHI materials purely due to atomic scale variations in the superconducting order parameter. This effect is present on AC edges but notably not on ZZ edges since there no microscopic gradients are generated.
In particular, it does not require any external tuning, either by applying a finite sublattice staggering or by creating macroscopic gradients, e.g., by creating a Josephson junction. We also note that all OT pair amplitudes are fully imaginary and thus they preserve time-reversal symmetry according to \req{eq:TRSconds}.

This OT state found at the AC edge is very different in both nature and origin from the OS state, the other odd-$\omega$ state already discussed. First of all, a finite OT amplitude requires an in-edge gradient of the conventional even-$\omega$ order parameter, which is only present on the AC edge, while the OS state exists also in the metallic bulk and generally at the ZZ edge. The varying gradient at the AC edge is also the origin of the alternating sign of the OT state along the edge as seen in Fig.~\ref{fig:ACOTS}(a). Secondly, the OT symmetry appears both in the $s$- and $s^+$-wave pair amplitudes, while the nature of the OS state makes it limited to the $s^+$ symmetry since it instead originates purely from the sublattice structure. 

\begin{figure}
\includegraphics[width=\linewidth]{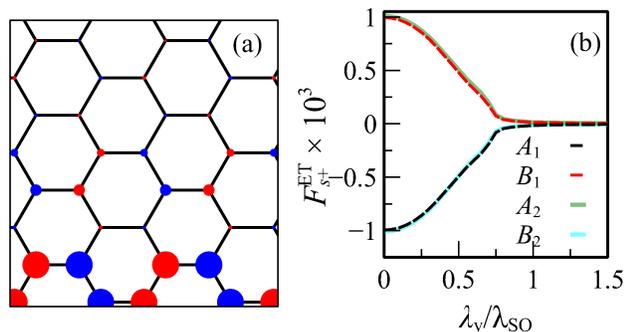}
\caption{\label{fig:ACETSp} ET NN bond $s^{+}$-wave pair amplitude for same parameters and geometry as in Fig.~\ref{fig:ACESS}.
}
\end{figure}

The OT NN $s^+$ state in Fig.~\ref{fig:ACOTSp} is naturally even under exchange of orbital index. But the question is if also a spin-triplet $s^+$-wave state that is odd in the orbital index is allowed. This state would then be an ET NN $s^+$-wave state. Such a state would be generated in the same way as the OS state appears out of the ES state in both ribbons and in the metallic bulk.  In Fig.~\ref{fig:ACETSp} we show that this state indeed exists. 
The evolution with $\stpar$ also clearly follows that of the OT, on-site $s$-wave pair amplitude in Fig.~\ref{fig:ACOTS}, which is exactly the same dependence the OS state inherits from the ES state.
We have thus for $s$-wave states found a complete reciprocity in oddness in the time and orbital domain for AC QSHI ribbons: there exist states that are odd (even) under both orbital and time exchange, which are all spin-singlet states, and there are also states that are odd (even) in time but even (odd) in orbital index, which are all spin-triplet states. The states that are odd in orbital symmetry  necessarily have a spatial $s^+$-wave symmetry, as such states can per definition not exist on-site. However, even orbital symmetric $s^+$-wave states are naturally accompanied by on-site $s$-wave states.

% ------------------------------------ %
% CONCLUSIONS:
% ------------------------------------ %
\section{\label{sec:Conclusions}Concluding remarks}
In summary, we have demonstrated that a buckled QSHI such as silicene, when proximity coupled to a conventional spin-singlet $s$-wave superconductor, exhibits multiple different odd-$\omega$ pair amplitudes. Even limiting the discussion to the disorder robust $s$-wave states we find, without fine-tuning, both odd-$\omega$ spin-triplet and spin-singlet states. 
Most universal is the OS state which appears both in the metallic bulk as well as in all insulating ribbons with only metallic edge states. This state is due to the intrinsic multi-orbital nature of the QSHI and is generated as soon as there is an asymmetry in the proximity-induced $s$-wave order parameter. Such an asymmetry is created in the metallic bulk by applying an electric field which causes a sublattice staggering. However, in ribbons with both ZZ and AC edges no staggering is needed since the required asymmetry is a natural consequence of the decay of superconductivity into the insulating bulk. Thus the OS states in QSHI ribbons are in an intrinsic state, requiring no fine tuning.

A recently discovered signature for odd-$\omega$ pairing from hybridization in multi-orbital systems is the appearance of extra hybridization gaps in the DOS at energies beyond the superconducting gap.~\cite{Komendova2015} These partial gaps appear at higher energies where two bands cross, one electron-like and one hole-like, in the Bogoliubov spectrum. However, we fail to observe such signatures in QSHIs. The reason for this is the peculiar band structure of the QSHI Hamiltonian $\mathcal{H}$. First of all, the Hamiltonian must have eigenvalues with opposite sign due to the particle-hole symmetry of the Bogoliubov spectrum. Furthermore, due to time-reversal symmetry, the eigenvalues at $-\bm{k}$ are the same as the eigenvalues at $\bm{k}$. 
Then, no matter the choice of the chemical potential $\mu$, a particle band at positive energy cannot hybridize with a hole band at negative energy, since the two are identically shifted by $2\mu$ in the QSHI Hamiltonian $\mathcal{H}$. This holds no matter the projection in $\bm{k}$ space, and, therefore, a particle-like and a hole-like band never cross either in the bulk case or for any edge orientation.    

Beyond the OS states, on the AC edges of QSHI ribbons we also find OT states. These are due to an effective gradient along the edge of the proximity-induced superconducting order. Externally enforced macroscopic gradients in QSHI and topological insulator surface states have previously been found to generate OT pairing,~\cite{AMBS_TI2012} Here however, no external gradient is needed but the OT state appears in AC ribbons simply due to atomic scale gradients. These gradients are present due to a combination of the AC edge geometry and the decay of the edge state and thus superconductivity into the bulk of the ribbon. The presence of this OT state also generates an ET $s^+$-wave state, and by exactly the same mechanism the OS state appears due to the ES state. Thus AC edges of QSHI ribbons give rise to all possible combinations of superconductivity--ES, OS, OT, and ET--with still intact disorder robust $s$-wave spatial parity. 

In combination our results show that two unrelated mechanisms generate two very different odd-$\omega$ superconducting pairing states in QSHIs, such as silicene, when in proximity to an external conventional superconductor. This makes superconducting QSHIs a very exciting playground for studying not only the existence of odd-$\omega$ superconductivity, but also the interplay between different odd-$\omega$ states.

\begin{acknowledgments}
We thank A.~V.~Balatsky and L.~Komendova for discussions related to previous work and acknowledge financial support from the Carl Trygger's Foundation, the Swedish Research Council (Vetenskapsr\aa det) Grant No.~621-2014-3721, and the Wallenberg Academy Fellows program of the Knut and Alice Wallenberg Foundation.
\end{acknowledgments}

\bibliographystyle{apsrev4-1}
\bibliography{OddFreqSilicene_Ref2.bbl}
%\bibliography{OddwSC_Si}
\end{document}